\documentstyle[11pt,epsf,twoside]{article}
\newcommand{\lesssim}{~\mbox{\raisebox{0.45ex}{$<$}\hspace{-11pt}
\raisebox{-0.45ex}{$\scriptstyle{\sim}$}}~}

\newcommand{\gsim}{~\mbox{\raisebox{0.45ex}{$>$}\hspace{-11pt}
\raisebox{-0.45ex}{$\scriptstyle{\sim}$}}~}

\begin{document}
\title{ Higgs studies in ACFA Linear Collider Working Group}
\author{
  Shingo Kiyoura$^{1,2,\dagger}$\thanks{e-mail address:
    shingo.kiyoura@kek.jp; This work was supported in part by
    Grant-in-Aid for Scientific Research ${}^\dagger$(A)(1)(No.13047101),
    ${}^{\ddagger}$(C)(No. 13640309).},
  Shinya Kanemura$^1$,
  Kosuke Odagiri$^1$, \\
  Yasuhiro Okada$^{1,\ddagger}$,
  Eibun Senaha$^{3}$,
  Satoru Yamashita$^{4}$, 
  Yoshiaki Yasui$^1$ \\ \\
  $^1$ {\it KEK, 1-1 Oho, Tsukuba, Ibaraki 305-0801, Japan} \\
  $^2$ {\it Department of Radiological Sciences, Ibaraki
        Prefectural University} \\
       {\it of Health Sciences, Ami, Inashiki, Ibaraki 300-0394,
         Japan} \\
  $^3$ {\it Department of Particle and Nuclear Physics, the
        Graduate University} \\
       {\it for Advanced Studies, Tsukuba, Ibaraki 305-0801, Japan} \\
  $^4$ {\it ICEP, University of Tokyo, 7-3-1 Hongo, Bunkyo-ku,
          Tokyo 113-0033, Japan}
}
\date{}
\maketitle
\vspace{-10cm}
\begin{flushright}
KEK-TH-864    \\
January 2003  \\
\end{flushright}
\vspace{7.5cm}
\vspace{-0.7cm}
\begin{abstract}
We report the important topics in ACFA report\cite{ACFA} as well as
the recent progress in the ACFA Higgs working group.
\end{abstract}
\vspace{-0.7cm}
\section{Introduction}  \vspace{-0.3cm}
While the $SU(2)\times U(1)$ gauge interaction has been
precisely tested at LEP and SLC, we do not understand the mechanism of
the gauge-symmetry breaking.
The existence of a Higgs boson is expected to be discovered at
TEVATRON or LHC.
One of the main goals at a future Linear Collider (LC) is a precision
study on the Higgs sector.
At the machine with the center of mass energy ($\sqrt{s}$) of
300-500 GeV and an integrated luminosity (${\cal L}_I$) of
500 fb$^{-1}$, O($10^5$) Higgs bosons can be produced if the Higgs mass
is smaller than 200~GeV.
By measuring the Higgs couplings to gauge bosons and fermions as well
as the Higgs-self coupling, we could reveal the structure of the Higgs
sector.
In addition, a TeV-scale LC may enable us to explore the heavy Higgs
bosons beyond the standard model (SM).
In this talk, we report results of recent activity in the ACFA Higgs
Working Group.
\vspace{-0.5cm}
\section{Light Higgs Boson} \vspace{-0.3cm}
\subsection{Mass} \vspace{-0.2cm}
The mass of the Higgs boson (h) is a key parameter in the Higgs
sector.
Assuming that the SM is valid up to the Planck scale
($\Lambda=$10$^{19}$GeV), the mass of the Higgs boson is theoretically
constrained in the range between 135 GeV and 180GeV.
In the two Higgs doublet model (THDM), the mass of the lightest
CP-even Higgs boson ($m_h$) is expected in the region between 100 GeV
and 180 GeV for $\Lambda=10^{19}$ GeV in the decoupling regime where
only one Higgs boson is light \cite{THDM-mass-bound}.
In the minimal supersymmetric standard model (MSSM), we can derive 
the upper bound of $m_h$ without reference to the cut-off scale.
The lightest Higgs mass receives significant radiative corrections due
to the top quark and stops\cite{MSSM-Higgs-Mass} and detail studies
show $m_h \lesssim130$GeV.
For the extended SUSY model with the gauge singlet field (NMSSM),
the bound is about 150GeV if we assume that the theory is valid up to
the GUT scale\cite{singlet,minimal-x-section}.
The maximal Higgs mass corresponds to a lower $\tan\beta$ value, which is
quite different from the MSSM case where the Higgs mass bound
increases as $\tan\beta$ grows. \vspace{-0.4cm}
\subsection{Detectability} \vspace{-0.2cm}
The detection of the light Higgs boson in the SM requires only a few
fb${}^{-1}$\cite{ACFA}.
In the MSSM, we can detect at least one of two CP-even Higgs bosons
through $e^+ e^- \rightarrow hZ$.
Furthermore, the discovery of at least one of three CP-even Higgs
bosons is guaranteed in the
NMSSM\cite{minimal-x-section}. \vspace{-0.4cm}
\subsection{Model-Independent Analysis for the Higgs Boson Couplings}
\vspace{-0.2cm}
In order to distinguish various models, we introduce 
a model-independent parameterization for various couplings to the Higgs
boson as follows\cite{ACFA}, \vspace{-0.05cm}
\begin{eqnarray*}
  {\cal L} =   x \frac{m_b}{v}   h\overline{b}b
               + y (   \frac{m_t}v h\overline{t}t
                     + \frac{m_c}v h\overline{c}c)
             + z \frac{m_\tau}v  h\overline{\tau}\tau
               + u (g m_W^{} h W_\mu W^\mu 
                   + \frac{g_Z^{}}{2} m_Z^{} h Z_\mu Z^\mu),
\end{eqnarray*}
where the following four parameters, $x$, $y$, $z$, and $u$,
represent the multiplicative factors in the Higgs-boson coupling
constants with down-type quarks, up-type quarks, charged-leptons and
gauge bosons.
The SM corresponds to $x$ = $y$ = $z$ = $u$ = $1$.
This expression is valid for the SM, MSSM, NMSSM, and
multi-Higgs-doublet models without tree-level flavor changing neutral
current.
In Fig.1 we show the expected accuracy of the parameter
determination at JLC with $\sqrt{s}$=300GeV and ${\cal
  L}_I=500$fb$^{-1}$ for $m_h=120$GeV.
The reference point is taken to be the SM case.
We use the measurement accuracy listed in Table 2.7 of
Ref.\cite{ACFA}.
We can see that $u$ and $x$ parameters are determined to a few \%
level, and $y$ and $z$ are constrained to
less than 10\%.
In the figure, we also show points corresponding to several input
parameters in the MSSM. From the correlation of the four parameters
determined at the LC experiment, it is possible to distinguish various
models.
For example, Type-I THDM which predicts the relation
$x$=$y$ and the MSSM have different relation in $x$-$y$ space.
For a large $\tan\beta$ value, the allowed range of
the $x$-$z$ space can deviate from
the $x$=$z$ line for the MSSM because of the SUSY corrections to the
$hb\overline{b}$ vertex\cite{epsb},
\begin{center}
  \begin{minipage}[t]{16cm}
    \vspace{-1.5cm}
    \hspace{-1cm}
      \epsfxsize=5.2cm
      \epsfbox{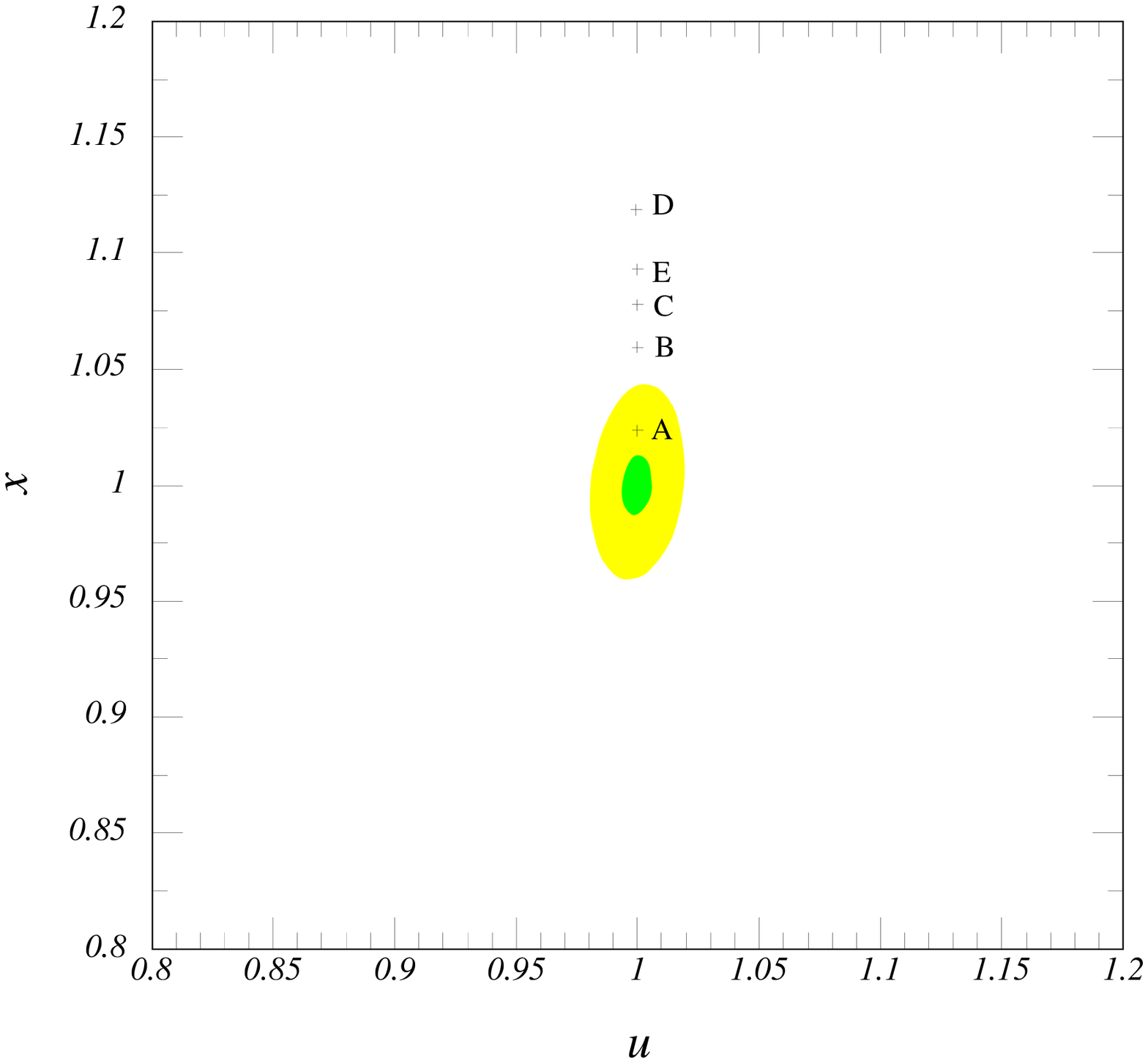}
      \epsfxsize=5.2cm
      \epsfbox{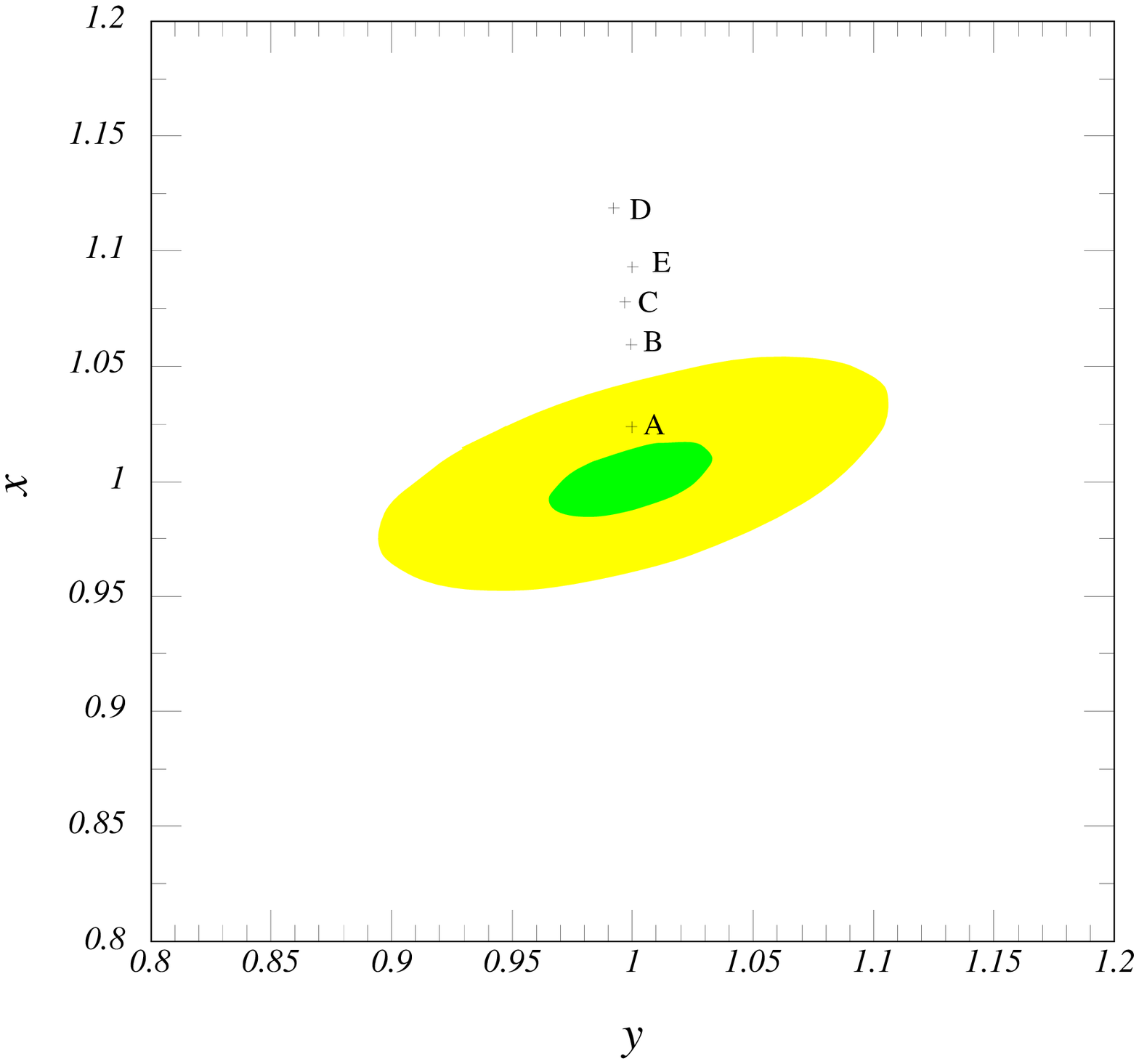}
      \epsfxsize=5.2cm
      \epsfbox{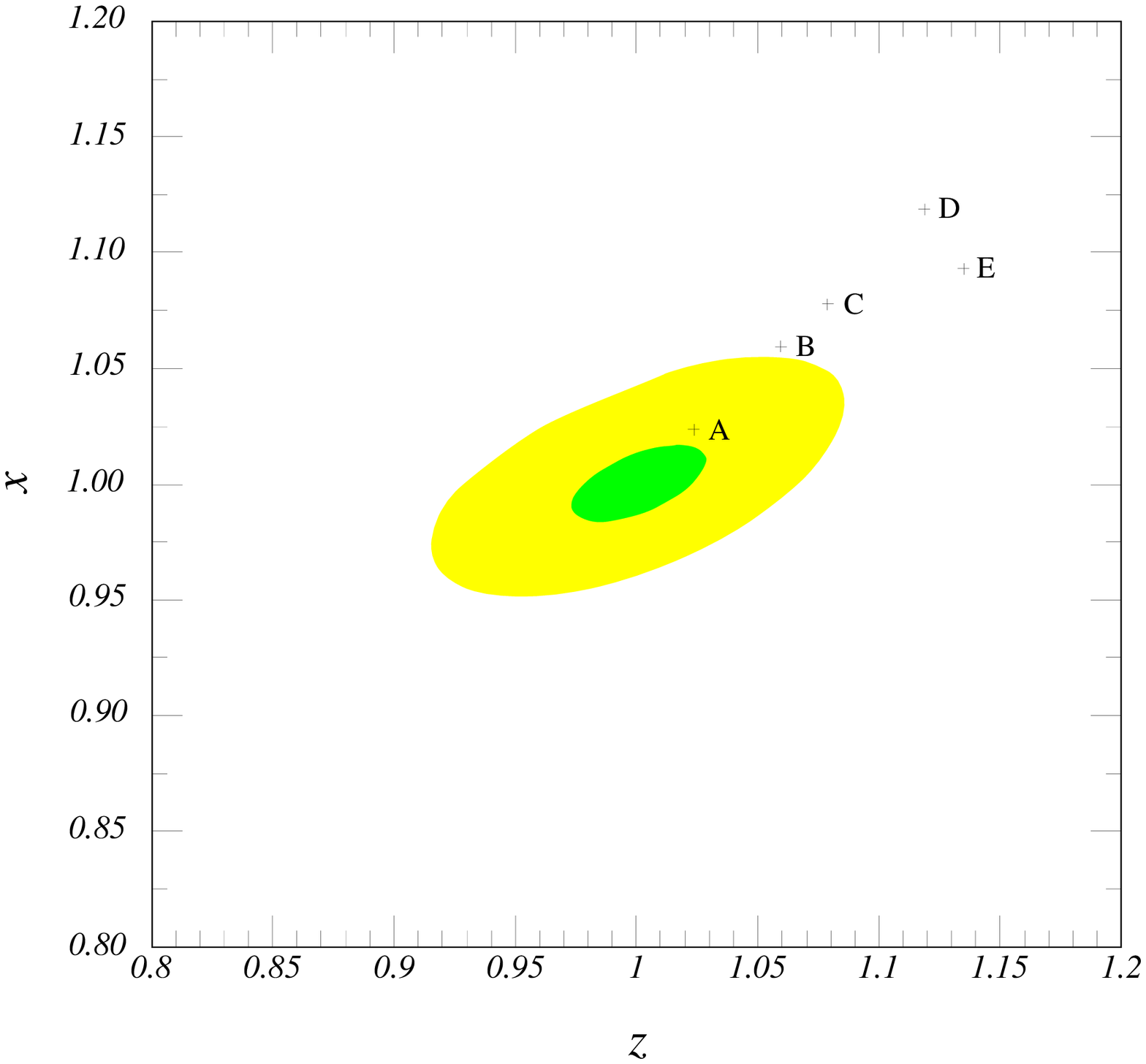}
  \end{minipage}
  \begin{minipage}[t]{14cm}
    \small Fig.1:
    The inner (outer) contour is the 1 $\sigma$ (95\% CL)
    curve. The points A-E correspond to the projections on each
    plane of the $x,y,z$, and  $u$ values evaluated in the following
    parameter sets of the MSSM. ($M_A$ (GeV), $M_S$ (GeV),
    $A_t/M_S$, $\tan\beta$) are (1000,500,0,10), (600,550,-2,10),
    (550,1430,2,5), (450,25430,0,4) and (500,4810,2,30) for A-E,
    respectively. We also take the gluino mass and the higgsino mass
    parameter as 300GeV.
  \end{minipage}
\end{center}
as shown for the point E.
A detailed analysis of the general $hZV$ couplings, with $V=\gamma$ or
$Z$, is presented in Ref.\cite{HZV}.\vspace{-0.5cm}
\subsection{Indirect Determination of Heavy Higgs Boson Mass in MSSM}
\vspace{-0.2cm}
We discuss the extraction of the heavy Higgs mass from the branching
ratios of the lightest Higgs boson in the MSSM\cite{MA0}.
Within the approximation that the stop mixing is neglected in the
one-loop Higgs potential and the $hb\overline{b}$ vertex correction
is small, four double ratios of the Higgs branching ratios, ($Br(h
\rightarrow c\overline{c}) + Br(h \rightarrow gg))/Br(h
\rightarrow b\overline{b}$), ($Br(h \rightarrow c\overline{c}) + Br(h
\rightarrow gg))/Br(h \rightarrow \tau^+ \tau^-$), $Br(h \rightarrow
W^{(*)}W^{(*)})/Br(h \rightarrow b\overline{b}$), and $Br(h
\rightarrow W^{(*)}W^{(*)})/Br(h \rightarrow \tau^+ \tau^-)$, are
approximately given by $R(m_A)\equiv(m_A^2-m_h^2)^2/(m_A^2+m_Z^2)^2$,
for $m_A>200$GeV and $\tan\beta\gsim 2$, where $m_A$ is the mass of
the CP-odd Higgs boson. Since $R(m_A)$ depends only on $m_A$ once the
lightest Higgs mass is measured, the double ratios of the branching
ratios are useful to constrain the mass of the heavy Higgs
boson.
In Fig.2, we show the precision of the indirect determination
on $m_A$ from the above branching ratios at JLC with ${\cal L}_I=500$
fb${}^{-1}$ at $\sqrt{s}=300$GeV for $m_h=120$GeV.
The theoretical uncertainty of the branching ratio calculation in the
SM and the estimated experimental statistical errors are summarized in
Table 2.9 of Ref.\cite{ACFA}.
The combined error to determine $m_A$ from ($Br(h \rightarrow
c\overline{c}) + Br(h \rightarrow gg))/Br(h \rightarrow \tau^+\tau^-)$
and $Br(h \rightarrow W^{(*)}W^{(*)})/Br(h \rightarrow \tau^+\tau^-)$
is 5.3\%.
In order to draw the constraints on $m_A$, we use the 5.3\% error and
assume that these ratios normalized by the SM values are given by
$R(M_A)$\footnote{However, in the presence of the stop mixing, the
SUSY corrections on the relation between the double ratios of the
Higgs branching ratios and $R(m_A)$ can be significant for
$\tan\beta>30$ or large $|\mu|\simeq$ 2TeV even if $\tan\beta$ is
not large\cite{MA1}.}.
In the figure, we can see that accuracy of the $m_A$
\begin{minipage}[h]{14cm}
  \begin{minipage}{4.8cm}
    \begin{center}
      \epsfxsize=4.7cm
      \epsfbox{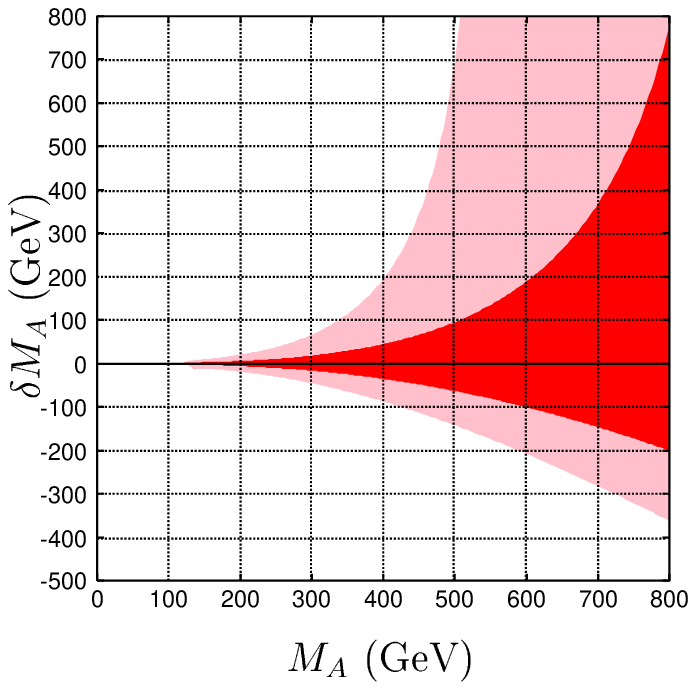}
    \end{center}
  \end{minipage}
  \
  \begin{minipage}[h]{8.4cm}
    {\small Fig.2: Accuracy of the $M_A$ determination as a
    function of $M_A$ from branching ratio measurements. The dark area
    corresponds to the error of $M_A$ from $Br(h \rightarrow
    c\overline{c}) + Br(h \rightarrow gg)$, $Br(h \rightarrow \tau^+
    \tau^-)$, and $Br(h \rightarrow W^{(*)}W^{(*)})$ measurements at
    JLC. The light area is obtained by the assumption that
    $\Gamma(h\rightarrow W^{(*)}W^{(*)})/\Gamma(h\rightarrow \tau^+
    \tau^-)$ is determined in 15\% accuracy, which corresponds to an
    estimated statistical error at LHC.}
  \end{minipage}
\end{minipage}
\\ \\
determination (dark area) is about $\pm100$ GeV for $m_A$ = 500 GeV.
This is compared with the typical accuracy of $m_A$ expected at LHC
from the measurement of the ratio $\Gamma(h \rightarrow
W^{(*)}W^{(*)})/\Gamma(h \rightarrow \tau^+\tau^-)$ (light area).
\vspace{-0.4cm}
\subsection{Top Yukawa Coupling Constant} \vspace{-0.2cm}
The top Yukawa coupling constant will be measured through the
$t\overline{t}h$ production process.
In the SM the production cross section becomes maximal around $\sqrt{s}
= 700$ GeV for $m_h=120$ GeV.
The expected accuracy for the top Yukawa coupling in the SM is 4.2\%
at $\sqrt{s}=700$ GeV with ${\cal L}_I =500$ fb${}^{-1}$.
\vspace{-0.4cm}
\subsection{Higgs Self-coupling} \vspace{-0.2cm}
The measurement of the Higgs self-coupling is crucial to test
the Higgs mechanism.
The trilinear Higgs coupling can be measured via $e^+e^- \rightarrow
Zhh$ and $\nu\overline{\nu}hh$ if the Higgs boson is light\cite{hhh}.
We have started systematic studies on the self coupling measurement
for various $m_h$ and $\sqrt{s}$\cite{hhh1}.
At $\sqrt{s} = 500$GeV with ${\cal L}_I=1$ab${}^{-1}$,
the expected statistical error of the trilinear coupling is about 20\%
for the SM Higgs boson for $m_h\lesssim 150$GeV.
For $\sqrt{s}~\gsim 1$ TeV, due to the enhancement of the $W$-fusion
process, the accuracy is expected to be better than
10\%\cite{hhh1,hhh2}.
Accurate information on the self-coupling is important to discriminate
models beyond the SM.
Even when all the Higgs couplings except for the self-interactions
are in good agreement with the SM predictions, the Higgs
self-couplings can significantly deviate from the SM prediction due to
the non-decoupling quantum effects of heavy particles. In the THDM
case, the radiative corrections of O(100)\% on the self-coupling is 
possible\cite{hhh-THDM}.
%
\vspace{-0.4cm}
\section{Heavy Higgs Boson}
\vspace{-0.2cm}
\subsection{Discovery Contour at the LC} \vspace{-0.2cm}
Next, we discuss the discovery potential at the LC for the heavy Higgs
bosons in the MSSM.
In Fig.3, we show the cross-section contours in $(m_A,\tan\beta)$
plane for the
\begin{minipage}[t]{16cm}
    \hspace{-1cm}
    \epsfxsize=16cm
    \epsfbox{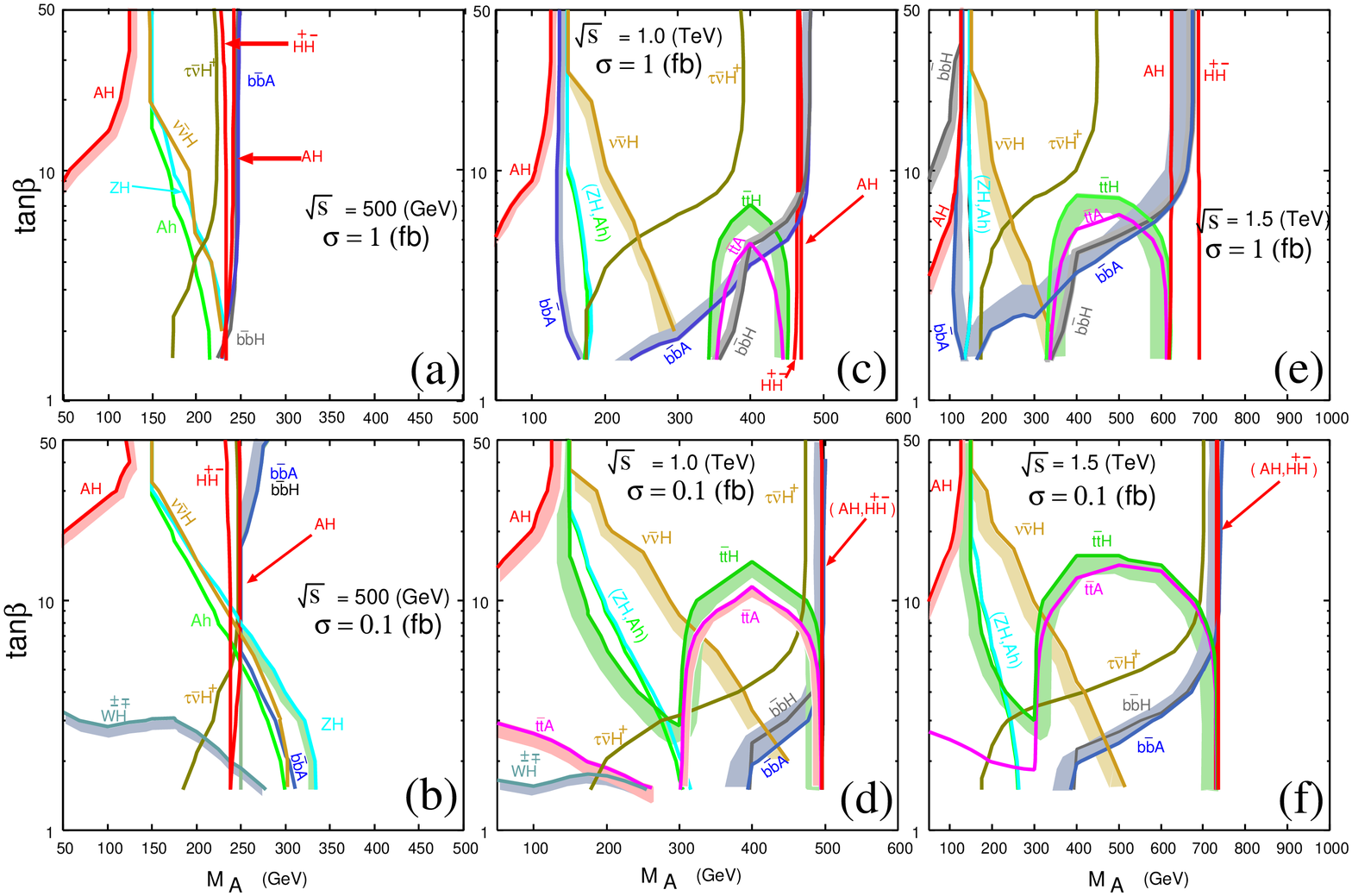}
\end{minipage}
\begin{minipage}{14cm}
  \begin{center}
    \small Fig.3 The discovery contours of the
    heavy Higgs bosons at the LC. See text.
  \end{center}
  \label{discovery-contour}
\end{minipage}
following processes:
$e^+e^- \rightarrow$ $ZH$, $Ah$, $AH$, $H^+H^-$, $W^{\pm}H^{\mp}$,
$b\overline{b}A$, $b\overline{b}H$, $t\overline{t}A$,
$t\overline{t}H$,
and $\nu\overline{\nu}H$.
We use GRACE/SUSY\cite{grace} to calculate the production cross-sections.
One-loop induced process of $W^+H^-$ was calculated as in 
Ref.\cite{single-charged0}.
Masses of the Higgs bosons and the mixing angle of the
neutral Higgs bosons are obtained by using
FeynHiggs\cite{feynhiggs},
where we assume the diagonal masses to be (1TeV)${}^2$ in the stop
mass matrix and the maximal stop mixing.
We adopt HDECAY\cite{hdecay} to calculate decay widths of the Higgs
bosons.
Fig.3 show the cross-section contours: (a) for
$\sqrt{s}$ = 500 GeV and $\sigma$ = 1 fb, (b) for $\sqrt{s}$ = 500 GeV
and $\sigma$ = 0.1 fb, (c) for $\sqrt{s}$ = 1.0 TeV and $\sigma$ =
1 fb, (d) for $\sqrt{s}$ = 1.0 TeV and $\sigma$ = 0.1 fb, (e) for
$\sqrt{s}$ = 1.5 TeV and $\sigma$ = 1 fb, and (f) for $\sqrt{s}$ = 1.5
TeV and $\sigma$ = 0.1 fb.
These contours can be translated to the discovery contours if
the sensitivity reach to those cross-sections.
In Fig.3 (a), (c), (d), (e) and (f),  the mass reach for $A$, $H$, and
$H^{\pm}$ at the LC is determined by half of $\sqrt{s}$.
In Fig.3(b), if the sensitivity reachs to 0.1fb,
the discovery contours for the $b\overline{b}A$ and $b\overline{b}H$
modes go beyond $\sqrt{s}/2$ for large $\tan\beta$.
For $\tan\beta\lesssim$10, the $ZH$ ($Ah$) mode is
available above $m_A > \sqrt{s}/2$.
In Fig.3(c)-(f), cross-section contours of $e^+e^-
\rightarrow t\overline{t}A$, $t\overline{t}H$, $b\overline{b}A$, and
$b\overline{b}H$ exhibit the dependence on $\tan\beta$ in
350GeV$\lesssim m_A\lesssim \sqrt{s}/2$.
These processes include $e^+e^- \rightarrow AH$ followed by $A$ or $H$
decaying into the $b\overline{b}$ or $t\overline{t}$ quark pair.
The $\tan\beta$ dependnce shown in Fig.3(c)-(f) is 
caused by the branching ratio of the heavy Higgs bosons.
Fig.3(c)-(e) also show that the LC will cover the region of moderate
$\tan\beta\lesssim10$ and $M_A\lesssim\sqrt{s}/2$ where the detection
of the heavy Higgs bosons at LHC is expected to be difficult.
If kinematically allowed, the heavy Higgs bosons are expected to be
found in several modes at the LC.
This would be useful in determining model parameters such as
$\tan\beta$ and $m_A$ in the MSSM, or in discriminating different
models from consistent determination of these parameters.
\vspace{-0.4cm}
\subsection{Single Charged Higgs Production}
\vspace{-0.2cm}
The charged Higgs pair production cross-section can be $10-100$ fb if
kinematically allowed.
Above the pair-production threshold, the single charged Higgs
  production is still possible\cite{single-charged1, single-charged3,
  single-charged2}.
For large $\tan\beta$, production cross-sections for the processes,
$e^+e^-\rightarrow $ $\tau^+ \nu H^-$ and $t\overline{b}H^-$, are
enhanced\cite{single-charged1,single-charged3}. For small $\tan\beta$,
the production cross-section of the $e^+e^-\rightarrow W^{\pm}H^{\mp}$
process becomes large due to the top and bottom quark
loops\cite{single-charged0}.
In addition the SUSY loop corrections to this mode was calculated in
Ref.\cite{single-charged2}.
\vspace{-0.4cm}
\subsection{Photon Collider}
\vspace{-0.2cm}
One of the important motivations for the
$\gamma \gamma$ option is to study the s-channel production of the
neutral Higgs bosons\cite{photon-higgs1}.
This provides the discovery potential for the MSSM heavy Higgs bosons
in moderate-$\tan\beta$ parameter space.
The kinematical reach will be extended to 0.8 $\sqrt{s_{ee}}$, where
$\sqrt{s_{ee}}$ is the center of mass energy for the $e^+e^-$
collider.
In addition, we can determine the CP parity of the heavy Higgs boson
through the process, $\gamma\gamma \rightarrow t\overline{t}$ by
measuring the helicity of the top quark\cite{photon-CP}.
The yield of a heavy charged Higgs boson at a $\gamma\gamma$ collider
is typically one order of magnitude larger than that at an $e^+e^-$
collider.
Moreover, a polarized $\gamma\gamma$ collider can determine the
chirality of the Yukawa couplings of fermions with charged Higgs boson
via single charged Higgs boson production and, thus, discriminate
models of new physics\cite{single-charged-photon}. \vspace{-0.5cm}

\end{document}